\documentclass[letterpaper, 10 pt, conference]{ieeeconf}  

\IEEEoverridecommandlockouts                              

\makeatletter

\let\proof\@undefined
\let\endproof\@undefined
\makeatother

\title{\LARGE \bf
Data-Driven Learning of Safety-Critical Control \\with Stochastic Control Barrier Functions
}

\author{Chuanzheng Wang$^\dagger$, Yiming Meng$^\dagger$, Stephen L. Smith, Jun Liu
\thanks{$^\dagger$Equal contribution}
\thanks{Chuanzheng Wang, Yiming Meng and Jun Liu are with the Department of Applied Mathematics,        
University of Waterloo, Waterloo, Ontario, Canada, 
        {\tt\small \{cz.wang, yiming.meng, j.liu\}@uwaterloo.ca}}%
\thanks{Stephen L. Smith is with the Department of Electrical and Computer Engineering, 
University of Waterloo, Waterloo, Ontario, Canada, 
        {\tt\small stephen.smith@uwaterloo.ca}}%
}

\usepackage{amsmath,amssymb,amsthm,amstext}
\usepackage{cite}
 
 \usepackage{algorithm,algorithmic}

 \usepackage{graphicx}
\usepackage{subcaption}
 \usepackage{mwe}

 \newcommand{\A}{\mathcal{A}}

  \newcommand{\eps}{\varepsilon}

 \newcommand{\ub}{\mathbf{u}}

 \newtheorem{definition}{Definition}[section]
\newtheorem{definitionA}{Definition A.\ignorespaces}
 \newtheorem{remark}[definition]{Remark}
 \newtheorem{prop}[definition]{Proposition}
 
 \newtheorem{thmA}{Theorem A.\ignorespaces}

 \newtheorem{ass}[definition]{Assumption}
 \newtheorem{prob}[definition]{Problem}
\usepackage{xcolor}

\usepackage{booktabs}
\usepackage{mathrsfs}
\usepackage{dsfont}
\newcommand{\R}{\mathbb{R}}
\newcommand{\pp}{\mathbb{P}}
\newcommand{\ee}{\mathbb{E}}
\newcommand{\ff}{\mathcal{F}}
\newcommand{\ppp}{\mathbf{P}}
\newcommand{\eee}{\mathbf{E}}
\newcommand{\U}{\mathcal{U}}
\newcommand{\X}{\mathcal{X}}

\newcommand{\Aa}{\mathcal{A}}
\newcommand{\ex}{\tau_\text{ex}}
\newcommand{\sig}{\varsigma}
\newcommand{\cj}{\wedge}

\begin{document}

\maketitle
\thispagestyle{empty}
\pagestyle{empty}


\begin{abstract}
 Control barrier functions are widely used to synthesize safety-critical controls. The existence of Gaussian-type noise may lead to unsafe actions and result in severe consequences. While studies are widely done in safety-critical control for stochastic systems, in many real-world applications, we do not have the knowledge of the stochastic component of the dynamics. In this paper, we study safety-critical control of stochastic systems with an unknown diffusion part and propose a data-driven method to handle these scenarios. More specifically, we propose a data-driven stochastic control barrier function (DDSCBF) framework and use supervised learning to learn the unknown stochastic dynamics via the DDSCBF scheme. Under some reasonable assumptions, we provide guarantees that the DDSCBF scheme can approximate the It\^{o} derivative of the stochastic control barrier function (SCBF) under partially unknown dynamics using the universal approximation theorem. We also show that we can achieve the same safety guarantee using the DDSCBF scheme as with SCBF in previous work without requiring the knowledge of stochastic dynamics. We use two non-linear stochastic systems to validate our theory in simulations. 
\end{abstract}

\section{INTRODUCTION}\label{sec:intro}
Safety-critical control is playing an important role in real-world control problems. In these applications, one must solve a control problem not only to achieve control performance objectives, but also to provide control actions with guaranteed safety \cite{garcia2015comprehensive}. Safety-critical control arises in a variety of practical applications including industrial robotics, medical robotics as well as self-driving vehicles. Since the notion of safety-critical control was first introduced in \cite{lamport1977proving}, there has been extensive research in safety verification problems, e.g., using discrete approximations \cite{ratschan2007safety} and computation of reachable sets \cite{girard2006efficient}.

Recently, control barrier functions (CBFs) have been widely used to deal with safety-critical control \cite{ames2019control}. Quadratic programming (QP) problems are used to solve safety-critical control with constraints from CBFs, together with control Lyapunov functions (CFLs) for achieving stability objectives \cite{ames2014control}. The authors show that the safety criteria can be transformed into linear constraints of the QP problems. By taking the derivative of the CBF, the control inputs can be treated as the decision variables of the QP problem so that we can find a sequence of actions that can guarantee safe trajectories. The authors in \cite{rauscher2016constrained} then show that finding safe control inputs by solving QP problems can be extended to an arbitrary number of constraints and any nominal control law. Consequently, safety-critical control using QP with CBF conditions have been applied in a wide range of applications such as lane keeping \cite{ames2016control} and obstacle avoidance \cite{chen2017obstacle}. However, for many applications in robotics \cite{hsu2015control}, the derivative of the CBFs is not dependent on the control input, and thus the control action can not be directly solved using a QP. To address this issue, CBFs have been extended to exponential control barrier functions (ECBF) to handle high relative degree constraints using input-output linearization \cite{nguyen2016exponential,xiao2019control}.

On the other hand, models used to design controllers are imperfect and this imperfection may lead to dangerous behavior. Consequently, designing controllers considering uncertainty is important in practical applications. In \cite{taylor2020learning} and \cite{wang2021learning}, a bounded disturbance is considered for the model, in which the time derivative of the barrier function is separated into the time derivative of the nominal barrier function and a remainder that can be approximated using neural networks. For systems driven by Gaussian-type  noise, stochastic differential equations (SDEs) are usually used to characterize the effect of randomness. Studies for stochastic stability of diffusion-type stochastic differential equations have seen a variety of applications in verifying probabilistic quantification of safe set invariance \cite{kushner1967stochastic}. Control barrier functions for stochastic systems have also been studied in recent years. The authors in \cite{sarkar2020high} applied the strong set-invariance certificate from \cite{clark2019control} to high-order stochastic control systems using stochastic reciprocal control barrier functions (SRCBFs) and \cite{santoyo2021barrier} investigates the worst-case safety verification utilizing stochastic zeroing control barrier functions (SZCBFs) regardless of the magnitude of noise. In \cite{wang2021safety}, the authors proposed the stochastic control barrier function (SCBF) with milder conditions at the cost of sacrificing the almost sure safety. 

However, in some practical scenarios, we do not have precise information about the Brownian motion that is affecting the system. In this case we cannot calculate the generator of the control barrier functions that is used in the corresponding QP problem. In \cite{nejati2021data}, the authors propose a method of estimating the value of the generator of a given function at a specific point within the domain. In this paper, we extend this idea to the whole state space and use a data-driven method to approximate the generator of the control barrier function globally. We propose a data-driven stochastic control barrier function (DDSCBF) framework for controlling stochastic systems with unknown diffusion parts. We show that under some reasonable assumptions, the DDSCBF scheme can approximate the It\^{o} derivative of a stochastic control barrier function (SCBF) as in our previous work and we can achieve the same safety probability as with SCBF in previous work \cite{wang2021safety}. We also validate our approach using two non-linear SDEs. 

\textbf{Notation:} 
We denote the $n$-dimensional Euclidean space by $\R^n$.  We denote $\R$ the set of real numbers, and $\R_{\geq 0}$ the set of nonnegative real numbers.  Given $a,b\in\R$, we define $a\wedge b:=\min(a,b)$. 
Let $C_b(\cdot)$ be the space of all bounded continuous functions/functionals $f: (\cdot)\rightarrow\R$. A  continuous and strictly increasing function $\alpha:\,
\R_{\ge 0}\rightarrow\R_{\ge 0}$ is said to belong to class $\mathcal{K}$ if $\alpha(0)=0$. 

For a given set $A\subseteq\R^n$, we denote by $A^c$ the complement of the set $A$ (i.e.,  $\R^n\setminus A$); denote by $\bar{A}$ 
(resp. $\partial A$) the closure (resp. boundary) of $A$.  


For any 
stochastic processes $\{X_t\}_{t\geq 0}$ we use the shorthand notation $X:=\{X_t\}_{t\geq 0}$. For controlled process under some $u$, we use $X^u$ in short for $\{X_t^u\}_{t\geq 0}$. For any stopped process $\{X_{t\cj\tau}\}_{t\geq 0}$, where $\tau$ is a stopping time, we use the shorthand notation $X^\tau$. We denote the Borel $\sigma$-algebra of a set by $\mathscr{B}(\cdot)$.

\section{Preliminaries and Problem Definition} 
\subsection{System Description}
Given 
a state space $\mathcal{X}\subseteq\R^n$ and a (compact) set of control values $\U\subset \R^p$, consider a 
continuous-time stochastic dynamical system 
\begin{equation}\label{E:sys}
dX_t=(f(X_t)+g(X_t)u(t))dt+b(X_t)dW_t, \;X_0=x,
\end{equation}
where $u:\R_{\geq 0}\rightarrow\U$ is a bounded measurable control signal; $W$ represents a $d$-dimensional standard Wiener process; $f:\X\rightarrow \R^n$ is a locally Lipschitz non-linear vector field; $g:\X\rightarrow\R^{n\times p}$ and $b:\X\rightarrow\R^{n\times d}$ are smooth mappings.

Note that from a modelling point of view, we usually do not specify in \textit{a priori} a Wiener process \cite{oksendal2003stochastic}. In addition, for the purpose of verifying dynamical behaviors in probability laws, it is not necessary to restrict ourselves to a specified probability space.  We consider the following natural sense of solution concept. 

\begin{definition}[Weak solutions]
For each fixed signal $u$, the system \eqref{E:sys} admits a weak solution if there exists a filtered probability space $(\Omega^\dagger,\mathscr{F}^\dagger,\{\mathscr{F}^\dagger_t\},  \pp^\dagger)$, where a Wiener process $W$ is defined and a pair $(X^u,W)$ is adapted, such that $X^u$ solves the SDE \eqref{E:sys}.
\end{definition} 

Data-driven methods allow us to collect information of solutions only in the state space. However,  we are unclear about the base probability space $(\Omega^\dagger,\mathscr{F}^\dagger, \pp^\dagger)$ where the Wiener process $W$ is defined. We transfer information to the canonical space in the following standard way, which gives us the convenience to study the probability law of the weak solutions and the probabilistic behavior in the state space.

Define $\Omega:=C([0,\infty);\R^n) $ with coordinate process $\mathfrak{X}_t(\omega):=\omega(t)$ for all $t\geq 0$ and all $\omega\in\Omega$. Define $\ff_t:=\sigma\{\mathfrak{X}_s,\;0\leq s\leq t\}$  for each $t\geq 0$, then  the smallest $\sigma$-algebra containing the sets in every $\ff_t$, i.e.  $\ff:=\bigvee_{t\geq 0}\ff_t$, turns out to be same as $\mathscr{B}( \Omega) $. 
For a weak solution $X^u$ of \eqref{E:sys} given a  valid $u$, the induced measure (law)  $\ppp^{x,u}$ on $\ff$ is such that $\ppp^{x,u}(A)=\pp^\dagger\circ (X^u)^{-1}(A)$ for every $A\in\mathscr{B}( \Omega)$.
The canonical probability space for $X^u$ is then $(\Omega,\ff,  \ppp^{x,u})$. We also denote $\eee^{x,u}$ by the associated expectation operator w.r.t. $\ppp^{x,u}$. When $g\equiv0$,  $\ppp^{x,u}$ and $\eee^{x,u}$ are simplified to $\ppp^{x}$ and $\eee^{x}$.

We also do not exclude explosive solutions\footnote{See \cite[Section 5.5]{barlow1988brownian} for details.} in general. As a matter of fact, for $g\equiv 0$, under the assumptions on $f$ and $b$, for any $x\in\mathcal{X}$, there exists a stopping time $\ex$ such that $\ppp^x[\ex>0]=1$ and 
a unique local weak solution $X$ of \eqref{E:sys} for all $t\in (0,\ex)$ such that $\ex=\infty$ (exists globally) or $\lim_{t\nearrow\ex}\|X_t\|=\infty$ (explodes within finite time).

\begin{definition} [Infinitesimal generator of $X$]\label{def:gen}
Consider a fixed signal $u$. 
Let $X^u$ be the weak solution to \eqref{E:sys}. The infinitesimal generator $\Aa$ of $X^u$ is such that
\begin{equation}
    \Aa h(x)=\lim\limits_{t\downarrow 0}\frac{\eee^{x,u}[h(X_t^u)]-h(x)}{t},\;\;x\in\mathcal{X},
\end{equation}
for all test functions $h\in C^2(\R^n)$ such that the limit exists at $x$.
\end{definition}

\begin{remark}
Note that the solutions $X$ of \eqref{E:sys} are diffusion processes. Given each valid test function $h\in C^2(\R^n)$ and control input $u\in\mathcal{U}$, 
\begin{equation}\label{E: ito}
    \Aa h(x)=\nabla h(x)\cdot(f(x)+g(x)u)+\frac{1}{2}\operatorname{tr}\left[(bb^T)(x)\cdot h_{xx}(x)\right]
\end{equation}
and $\Aa h$ is a continuous function.
\end{remark}

\subsection{Set Invariance and Control}

\begin{definition}[Control strategy]
A control strategy is a set-valued function from the state space $\mathcal{X}$ to a subset of control values in $\mathcal{U}$:
\begin{equation}
   \kappa:\X\rightarrow 2^{\U}. 
\end{equation}
\end{definition}

We use a boldface $\ub$ to indicate a set of constrained control signals. A special set of such signals is given by a (deterministic) control strategy as defined below.

\begin{definition}[State-dependent control]\label{E: control constraint}
We say that a control signal $u$ conforms to a control strategy $\kappa$ for (\ref{E:sys}), and write $u\in \ub_\kappa$, if 
\begin{equation}
    u(t) \in \kappa(X_t),\quad\forall t\geq 0,
\end{equation}
where $X$ satisfies (\ref{E:sys}) with $u$ as input. The set of all control signals that conform to $\kappa$ is denoted by $\ub_\kappa$. 
\end{definition}

For stochastic dynamical systems with controls as given in \eqref{E:sys}, we define a probabilistic set invariance property for the controlled processes.

\begin{definition}[Controlled probabilistic invariance]
Consider a system \eqref{E:sys}, a set of control signals $\ub$, and $p\in[0,1]$. A set $\mathcal{C}\subseteq\X$ is said to be controlled $p$-invariant under $\ub$ for system \eqref{E:sys} with a specified $x\in \mathcal{C}$ if, for all $u\in\ub$,
\begin{equation}
    \ppp^{x,u}[X_t^u\in\mathcal{C}, \;0\leq t< \infty]\geq p.
\end{equation}

\end{definition}

\begin{subsection}{Problem Formulation}
For the rest of this paper, we consider a safe set of the form
\begin{equation}\label{E: safeset}
    \mathcal{C}:=\{x\in\X: h(x)\geq 0\},
\end{equation}where $h\in C^2(\R^n)$. We also define the boundary and interior of $\mathcal{C}$ explicitly as below
\begin{equation}
    \partial\mathcal{C}:=\{x\in\X: h(x)= 0\},
\end{equation}
\begin{equation}
    \mathcal{C}^\circ:=\{x\in\X: h(x)> 0\}.
\end{equation}

The objective of this paper is to control the stochastic system (\ref{E:sys}) with an unknown diffusion term to stay inside the safe set. The problem is defined as follows.

\begin{prob}
Given system \eqref{E:sys} with the $b$ unknown, a compact set $\mathcal{C}\subseteq\X$ defined by \eqref{E: safeset}, a point $x\in\mathcal{C}^\circ$, and a $p\in[0,1]$, design a (deterministic) control strategy $\kappa$ such that under $\ub_\kappa$, the interior $\mathcal{C}^\circ$ is controlled $p$-invariant for the resulting  solutions to \eqref{E:sys}.
\end{prob}
\end{subsection}

\begin{remark}
In this paper, we assume we have full knowledge of the drift term of the system, i. e., $f$ and g in \eqref{E:sys}. For uncertainty in drift term, refer \cite{taylor2020learning, wang2021learning} for detail.
\end{remark}

\section{Worst-case Probabilistic Quantification via Stochastic Control Barrier Function }
There witnesses a surge of applications of control synthesis for probabilistic safety problems using stochastic control barrier functions. Two commonly-used types of  stochastic control barrier functions, reciprocal type (SRCBF)\cite{clark2019control} and zeroing type (SZCBF) \cite{santoyo2021barrier}, are investigated. The recent work \cite{wang2021safety} identified the pros and cons of SRCBF and SZCBF and  proposed a middle-ground type SCBF\footnote{We use this acronym for the specified notion in \cite[Definition III.6]{wang2021safety} rather than the general type of stochastic control barrier functions.} as in \cite[Definition III.6]{wang2021safety}. 

Note that the function $h$ in \eqref{E: safeset} is already a potential SCBF candidate. We need to further impose a  condition on the drift term of its It\^{o} derivative along the sample paths, i.e. $\Aa h$, to make it effective. 

Due to the lack of information of the diffusion term in \eqref{E:sys}, we are unable to capture the correction term in the It\^{o} derivative of the nominal barrier function $h$ along sample paths. In other words, the second term in  \eqref{E: ito} is unknown. We use a data-driven method to approximate the function $\Aa h$, and impose similar barrier conditions on the approximated $\hat{\Aa}h$ for safety-critical control. Based on the partial observation of data, we show in this section that a degree of robustness in the barrier condition is necessary to balance the inaccuracy of data. A similar approach can be applied to derive the robustness for the other types of stochastic control barrier functions.

We suppose that data is sampled without control inputs. Then for each $x$, the law $\ppp^x$ process $X$ is independent of $u$. We further define the stopping time
$$\tau:=\inf\{t\geq 0: X_t\in \partial \mathcal{C}\}$$
for each sampled process. Let $\mathfrak{C}$ denote a finite subset of $\mathcal{C}$.

\subsection{Probability Estimation based on Partially Observed Data}

We make the following assumptions for the rest of derivation. We show in the next subsection that the assumptions are feasible for compact $\mathcal{C}$.

\begin{ass}\label{ass: learn}Let $\hat{\Aa}h$ be the approximation of $\Aa h$ based on the training set $\mathfrak{C}$. 
We assume that 
\begin{enumerate}
    \item[(i)] For any  $y\in\mathcal{C}$ and any $\eps>0$, 
    there exists an $x\in\mathfrak{C}$ such that\footnote{Note that $\tau<\ex$ with probability 1.}
    \begin{equation}\label{E: close}
    \begin{split}
     &\eee^{y,u}\sup_{t\in[0,\tau]}|\hat{\Aa}h(X^u_{t})-\Aa h(X^u_{t})|\\
     \leq & \eee^{x,u}\sup_{t\in[0,\tau]}|\hat{\Aa}h(X^u_{t})-\Aa h(X^u_{t})|+\eps.
    \end{split}
         \end{equation}
    \item[(ii)] For any $\sig\in(0,1]$, there exists a probability measure $\pp$ with marginals $\ppp^x$ for all $x\in\mathfrak{C}$ such that 
    \begin{equation}\label{E: prob}
        \ee\sup_{x\in\mathcal{C}}|\Aa h(x)-\hat{\Aa}h(x)|\leq\sig.
    \end{equation}
\end{enumerate}
Furthermore, we assume that both $\hat{\Aa}h$ and $\Aa h$ are Lipschitz continuous on the compact set $\mathcal{C}$. 
\end{ass}

We apply the approximated function $\hat{\Aa} h$ and show the worst-case safety probability 
of the controlled process under policy generated by the following robust scheme.

\begin{prop}\label{prop:guarantee}
Suppose we are given arbitrary $\sig>0, \eps>0$ and training set $\mathfrak{C}$. Let $\hat{\Aa}h$ be generated as in Assumption \ref{ass: learn}. Suppose that $\sup_{u\in\mathcal{U}}\hat{\Aa} h(x)\geq \sig+\eps$ for all $x\in \mathcal{C}$. 
Let $\upsilon(x)=\{u\in\mathcal{U}: \hat{\Aa} h(x)\geq \sig+\eps\}$. Then for any $x\in\mathcal{C}^\circ$ and $u\in\mathbf{u}_\upsilon$, we have
$$ \ppp^{x,u}[X_t^u\in\mathcal{C}^\circ,\;0\leq t<\infty]\geq \frac{h(x)}{\sup_{y\in\mathcal{C}}h(y)}$$
\end{prop}

\begin{proof}
Let $c=\sup_{y\in\mathcal{C}}h(y)$ and set $V=c-h$. Then for all $x\in\mathcal{C}^\circ$, we have $V(x)>0$ and $\hat{\Aa}V(x)\leq -(\sig+\eps)$. Note that
\begin{equation}\label{E: dyn}
\begin{split}
       \eee^{x,u}[V(X_{\tau\cj t}^u)]=V(x)+\eee^{x,u}\left[\int_0^{\tau\cj t}\Aa V(X_s^u)ds\right]
\end{split}
\end{equation}
and by assumption, 

\begin{equation}\label{E: int}
\begin{split}
&\eee^{x,u}\left[\int_0^{\tau\cj t}\Aa V(X_s^u)ds\right]\\
 = & \eee^{x,u}\left[\int_0^{\tau\cj t}\Aa V(X_s^u)-\hat{\Aa}V(X_s^u)\;ds\right]\\
 &+\eee^{x,u}\left[\int_0^{\tau\cj t}\hat{\Aa}V(X_s^u)\;ds\right]\\
 \leq & \int_0^{\tau\cj t}\eee^{x,u}|\Aa V(X_s^u)-\hat{\Aa}V(X_s^u)|\;ds-(\sig+\eps)\cdot(\tau\cj t)\\
 \leq & \int_0^{\tau\cj t}\ee \sup_{s\in[0,\tau]}|\Aa V(X_s^u)-\hat{\Aa}V(X_s^u)|\;ds-\sig\cdot(\tau\cj t)\\
        \leq & \int_0^{\tau\cj t}\ee\sup_{x\in\mathcal{C}}|\Aa V(x)-\hat{\Aa}V(x)|\;ds-\sig\cdot(\tau\cj t)\leq  0,\\
\end{split}
\end{equation}
where the fifth line of the above is to transfer information from arbitrary $x\in\mathcal{C}$ to the data used in $\mathfrak{C}$. The mismatch of measure provides an extra error of $\eps$. 
Hence, by \eqref{E: dyn}, we have
\begin{equation}\label{E: dyn2}
    \eee^{x,u}[V(X_{\tau\cj t}^u)]\leq V(x),\;\;\forall t\geq 0.
\end{equation}
On the other hand, for all $t\geq 0$,
\begin{equation}\label{E: geq}
    \begin{split}
        \eee^{x,u}[V(X_{\tau\cj t}^u)]&\geq \eee^{x,u}[\mathds{1}_{\{\tau\leq t\}}V(X_{\tau\cj t}^u)]\\
        & \geq\ppp^{x,u}[\tau\leq t]\cdot\eee^{x,u}[V(X^u(\tau)]\\
                       & >c\cdot\ppp^{x,u}[\tau\leq t].
    \end{split}
\end{equation}
Therefore, by \eqref{E: dyn2} and \eqref{E: geq}, we have
\begin{equation}
    \ppp^{x,u}[\tau\leq t]<\frac{V(x)}{c},\;\forall t\geq 0.
\end{equation}
Sending $t\rightarrow\infty$ we get  $\ppp^{x,u}[\tau<\infty]\leq \frac{V(x)}{c}$ for all $x\in\mathcal{C}^\circ$. Rearranging this we can obtain the conclusion. 
\end{proof}
\begin{remark}
Note that (ii) in Assumption \ref{ass: learn} indicates that the error of estimation should converge in $L_1$, and cannot be replaced by in probability in the sense that, for every $\sig$, there exists a $\delta=\delta(\sig)$ such that
$$
        \pp\left[\sup_{x\in\mathcal{C}}|\Aa h(x)-\hat{\Aa}h(x)|>\delta\right]<\sig.
$$
The latter is not sufficient to show the last line of \eqref{E: dyn} in general. 
\end{remark}

\subsection{Feasibility of Assumptions}
Note that for the compact set $\mathcal{C}$ and for sufficiently dense training data, the conditions in Assumption \ref{ass: learn} can be  satisfied theoretically. We will show that both (i) and (ii) of Assumption \ref{ass: learn} require the selection of the training data but separately. Before  proceeding to the explanation, we introduce the following concepts. 

\begin{definition}\label{def: weak}
\textbf{(Weak convergence of measures and processes):}
Given any separable metric space $(\mathcal{S},\rho)$, 
a sequence of probability measure $\{\ppp^n\}$ on $\mathscr{B}(\mathcal{S})$ is said to weakly converge to $\ppp$ on $\mathscr{B}(\mathcal{S})$, denoted by $\ppp^n\rightharpoonup\ppp$, if for all  $f\in C_b(\mathcal{S})$ we have
$\lim_{n\rightarrow\infty}\int_{\mathcal{S}}f\;d\ppp^n=\int_{\mathcal{S}}f\;d\ppp. $ 
A sequence $\{X^n\}$ of continuous processes $X^n$ with law $\ppp^n$ is said to weakly converge (on $[0,T]$) to a continuous process $X$ with law $\ppp$, denoted by $X^n\rightharpoonup X$,  if for all $f\in C_b(C([0,T];\R^n))$ we have 
$\lim_{n\rightarrow\infty}\eee^{n}[f(X^{n})]=\eee[f(X)].$
\end{definition}

The following proposition  demonstrates a compactness of weak solutions starting from a compact set in a weak sense as in Definition \ref{def: weak}. We provide the rephrased version based on \cite[Theorem 1]{kisielewicz2007stochastic} and \cite[Corollary 1.1, Chap 3]{kisielewicz2013stochastic} as follows.  A detailed explanation can be found in \cite{meng2022acc}.

\begin{prop}\label{thm: weak_compact}
Given any compact set $\mathcal{C}$ and its associated first-hitting time $\tau$, given any sequence of stopped weak solutions $\{(X^{n})^\tau\}_{n=1}^\infty$ with $X^{n}(0)=x_n$, there exists a subsequence $\{(X^{n_k})^\tau\}$ and a process $X$ with $X(0)=x$ such that $x_{n_k}\rightarrow x$ and $(X^{n_k})^{\tau}\rightharpoonup X^{\tau}$.
\end{prop}

\subsubsection{Justification of Assumption \ref{ass: learn}(i)}

We observe that for each $x$ in a compact set $\mathcal{C}$, for any fixed $T>0$, the quantity $\sup_{t\in[0,\tau\cj T]}|\Aa h(\cdot)-\hat{\Aa}h(\cdot)|$ is a bounded function on the canonical space generated by $\mathcal{C}$ with measure $\ppp^x$. 
In view of Definition \ref{def: weak} and Proposition \ref{thm: weak_compact}, the quantity $$\left\{\eee^x\sup_{t\in[0,\tau\cj T]}|\Aa h(X^u_{t})-\hat{\Aa}h(X^u_{t})|\right\}_{x\in\mathcal{C}}$$ forms a compact set (in the conventional sense). By the boundedness assumption on $\mathcal{C}$, we have $\tau<\infty$ $\ppp^x$-a.s. for every $x\in\mathcal{C}$. Therefore, sending $T$ to infinity, we still have the compactness for
$$\left\{\eee^x\sup_{t\in[0,\tau]}|\Aa h(X^u_{t})-\hat{\Aa}h(X^u_{t})|\right\}_{x\in\mathcal{C}}.$$
By choosing $\mathfrak{C}$ sufficiently dense in $\mathcal{C}$, for each given $\eps>0$, we are able to build the $\eps$-net with centers in $\mathfrak{C}$ such that for any arbitrary $y\in\mathcal{C}$, there exists an $x\in\mathfrak{C}$ such that $\Aa h-\hat{\Aa}h$  are weakly $\eps$-close to each other in the sense of \eqref{E: close}. 

We then verify the feasibility of (ii) of Assumption \ref{ass: learn}.

\subsubsection{Approximating $\Aa h$ over a finite set}

Note that, following the procedure as in \cite{nejati2021data}, we are able to approximate $\Aa h$ by some $\tilde{\Aa}h$ at one single point $x\in\R^n$ at a time, whose precision is measured under the corresponding probability \footnote{In \cite{nejati2021data}, the authors used $\pp$, but  in our context it is recast to be $\pp^x$. The uniqueness of $\pp^x$ is by Kolmogrov's extension theorem. 
}  $\pp^x:=\otimes_{i=1}^\infty\ppp^x$. However, to fit the assumption, we need the precision to be measured in $L_1$ sense. 

By  \cite[Theorem 6]{nejati2021data}, for each $x\in\R^n$, we can utilize Lipschitz continuity of $f,g,b$ and the relation
$$\tilde{\Aa}_1h(x)=\frac{\eee^x[h(X^u_{\tau_s})]-h(x)}{\tau_s} $$ at some  deterministic sampling time $\tau_{s}$ to obtain the first-step approximation
\begin{equation}\label{E: primary}
|\tilde{\Aa}_1h(x)-\Aa h(x)|\leq \delta,
\end{equation}
where $\delta=C_1\tau_s+C_2\sqrt{\tau_s} $, and $C_1,C_2>0$ are constants generated by Lipschitz continuity. The precision $\delta$ can be arbitrarily small. 

Since $\tilde{\Aa}_1h(x)$ has used $ \eee^x[h(X^u_{\tau_s})]$, the authors in \cite{nejati2021data} then applied the law of large numbers (LLN) to approximate $ \eee^x[h(X^u_{\tau_s})]$ by 
$\frac{1}{n}\sum_{i=1}^n h(X^{u,(i)}_{\tau_s}) $ with i.i.d. $h(X^{u,(i)}_{\tau_s})$ draw from $\ppp^x$ at the marginal time $\tau_s$. The approximation
$$\tilde{\Aa}h= \frac{\frac{1}{n}\sum_{i=1}^n h(X^{u,(i)}_{\tau_s})-h(x)}{\tau_s} $$ 
  creates errors in probability w.r.t. $\pp^x$ as in \cite[Theorem 12]{nejati2021data}, i.e., for each $\beta\in(0,1]$, there exists a $\tilde{\delta}$ such that
$$\pp^x [|\Aa h(x)-\tilde{\Aa}h(x)|\leq \tilde{\delta}]>1-\beta. $$ 
Note that the only place that we introduce $\pp^x$ is when we use LLN. We need to leverage the convergence in the $L_1$ sense, i.e.,
\begin{equation}\label{E: l1}
    \ee^x\left|\frac{1}{n}\sum_{i=1}^n h(X^{u,(i)}_{\tau_s})-\eee^x[h(X^{u}_{\tau_s})]\right|\rightarrow 0.
\end{equation}
This is indeed the case as an existing result, even though it is seldom mentioned. Combining \eqref{E: l1} and \eqref{E: primary}, we can easily obtain that for each $x\in\R^n$, for any $\delta>0$, there exists a sufficiently large $n$ such that
\begin{equation}
    \ee^x\left|\tilde{\Aa}h(x)-\Aa h(x))\right|\leq \delta.
\end{equation}
We provide the proof for the $L_1$ convergence of LLN in the Appendix. 

Repeating the same process for $x$ over a finite set $\mathfrak{C}$ gives 
\begin{equation}\label{E: prod1}
            \sup_{x\in\mathfrak{C}}\ee\left[|\Aa h(x)-\tilde{\Aa}h(x)|\right]\leq \delta,
\end{equation}
where $\ee$ is the associated expectation w.r.t. $\pp:=\otimes_{x\in\mathfrak{C}}\pp^x$.

\subsubsection{Optimization error}

For any $\eta>0$, we assume there exists an optimizer that can learn an approximation $\hat{\Aa}h$ based on data $\left\{\tilde{\Aa}h(x):\; x\in\mathfrak{C}\right\}$ such that 
\begin{equation}
    \label{eq:eta}
    \sup_{x\in\mathfrak{C}}|\hat{\Aa}h(x)-\tilde{\Aa} h(x)|<\eta.
\end{equation}

\subsubsection{Generalization error}

By continuity of $\hat{\Aa} h(x)$ and $\Aa h(x)$, there exists some $x^*\in \mathcal{C}$ such that 
$$
\sup_{x\in\mathcal{C}}|\hat{\Aa} h(x)-\Aa h(x)| = |\hat{\Aa} h(x^*)-\Aa h(x^*)|. 
$$
For any $\theta>0$, by choosing $\mathfrak{C}$ to be sufficiently dense in $\mathcal{C}$ and the Lipschitz continuity of $\hat{\Aa} h(x)$ and $\Aa h(x)$ on $\mathcal{C}$, there exists some $y\in \mathfrak{C}$ such that 
$$
|\hat{\Aa} h(x^*) - \hat{\Aa} h(y)| \le \theta,\quad |\Aa h(x^*) - \Aa h(y)| \le \theta. 
$$
It follows that 
\begin{equation*}
    \begin{split}
       &\ee\left[\sup_{x\in\mathcal{C}}|\hat{\Aa} h(x)-\Aa h(x)|\right]\\
       =&\ee\left[|\hat{\Aa} h(x^*)-\Aa h(x^*)|\right]\\
       =&\ee\left[|\hat{\Aa} h(y)-\Aa h(y)+\hat{\Aa} h(x^*)-\hat{\Aa} h(y)\right.\\
       &\,\qquad\qquad\qquad\qquad\left.+\Aa h(y)-\Aa h(x^*)|\right]\\
       \le &\ee \left[|\hat{\Aa} h(y)-\Aa h(y)|\right] + 2\theta \\
       =&\ee \left[|\hat{\Aa} h(y) -\tilde{\Aa}h(y) + \tilde{\Aa} h(y) -\Aa h(y)|\right] + 2\theta \\
       \le& \ee\left[\sup_{y\in \mathfrak{C}}|\hat{\Aa} h(y) -\tilde{\Aa}h(y)|\right] + \ee\left[|\tilde{\Aa} h(y) -\Aa h(y)|\right] + 2\theta \\
       \le & \eta + \sup_{y\in \mathfrak{C}} \ee\left[|\tilde{\Aa} h(y) -\Aa h(y)|\right] + 2\theta \\
       \le & \eta + \delta + 2\theta \le \sig,
    \end{split}
\end{equation*}
where $\sig$ is from Assumption \ref{ass: learn}(ii), provided that we choose $\eta$, $\delta$, and $\theta$ sufficiently small. 

\begin{remark}
The final $\mathfrak{C}$ should be chosen based on all of the above criteria such that (i) and (ii) of Assumption \ref{ass: learn} can both be satisfied. 
\end{remark}

\section{Data-driven Stochastic Control Barrier Function Scheme for Safety-Critical Control}
In this section, we describe how do we use supervised learning to implement the DDSCBF scheme and do safety-critical control for stochastic systems with an unknown diffusion part. We use a neural network to approximate the derivative of an SCBF. The detail of data collecting and training is explained below.

Given an SDE as in \eqref{E:sys}, we have
\begin{equation*}
    \A h(x)=L_fh(x)+L_gh(x)u+\frac{1}{2}\operatorname{tr}\left[(bb^T)(x)\cdot h_{xx}(x)\right].
\end{equation*}
Since the only unknown part in the SDE is $b$, the unknown part in $\A h(x)$ is only $\frac{1}{2}\operatorname{tr}\left[(bb^T)(x)\cdot h_{xx}(x)\right]$. As we can see, this term is a function of $x$ only, so we can define a function $\Delta(x)=\frac{1}{2}\operatorname{tr}\left[(bb^T)(x)\cdot h_{xx}(x)\right]$ and accordingly 
\begin{equation*}
    \Aa h(x)=L_fh(x)+L_gh(x)u+\Delta(x).
\end{equation*}
We use supervised learning to learn $\Delta(x)$ (and hence some $\hat{A}h(x)$) that approximates $\A h(x)$.

Next we describe how to obtain the training data. We use a sampling method to collect data in order to learn $\Delta(x)$. First we sample a set with $N$ initial points $\{x_1, x_2,..., x_N\}$. At the initial point $x_i$ for $i\in\{1,2,...,N\}$, we sample $n$ one-step  transitions and reach to the next stage $x_{ij}$ for $j\in\{1,2,...,n\}$. We reset the point back to $x_i$ after each one-step transition. According to \cite{nejati2021data}, $\tilde{\A}h(x_i)$ can be estimated numerically by 
\begin{equation}\label{E:est_Ah}
    \tilde{\A} h(x_i)=\frac{\frac{1}{n}\sum\limits_{j=0}^{n} h(x_{ij})-h(x_i)}{\Delta t}.
\end{equation}
As a result,  obtain $\tilde{\Delta}(x_i)$ by 
\begin{equation*}
    \tilde{\Delta}(x_i)=\tilde{\A}h(x_i)-(L_fh(x_i)+L_gh(x_i)u).
\end{equation*}
We then add $\{x_i, \tilde{\Delta}(x_i)\}$ into a data set $D$, constructing a data set $D$ is of dimension $N$. Next we use learning to fit the data set. The process of collecting training data is shown as in shown as in Algorithm~\ref{alg:alg}.

Once we have collected the data set $D$, we construct a neural network $\mathcal{N}(x)$ and specify a loss function $\mathcal{L}$ using minimum square error (MSE). We use supervised learning to find the parameters of the network such that the $\frac{1}{N}\sum_{i}^{N}\mathcal{L}(\mathcal{N}(x_i), \tilde{\Delta}(x_i))$ is minimized. This implies that the neural network $\mathcal{N}(x)$ will approximate the function $\Delta(x)$. So the derivative of the SCBF $\A h(x)$ will be approximated by $\hat{\Aa}h(x):=L_fh(x)+L_gh(x)u+\mathcal{N}(x)$. As a result, we can use this approximated derivative of SCBF as QP constraints to guarantee safety-critical control for stochastic systems with unknown diffusion part as in \cite{wang2021safety}. The overall theoretical analysis of guarantees is shown in \ref{prop:guarantee} under Assumption \ref{ass: learn}. 

\begin{algorithm}[ht]
	\caption{Data-driven learning algorithm of SCBF}
	\label{alg:alg}
	\begin{algorithmic}[1]
	\REQUIRE An SDE as in \eqref{E:sys}, the number of initial points $N$, the number of trajectories sampled at each initial point $n$, an empty data-set $D$, a time step $\Delta t$, an initial neural network $\mathcal{N}(x)$.
	\STATE Initialize neural network
	\STATE Sample $N$ initial points $\{x_1, x_2,..., x_N\}$
	\FOR{$i$ in $N$}
	\FOR{$j$ in $n$}
	\STATE Get $x_{ij}$ from $x_{i}$ using Euler-Maruyama method according to \eqref{E:sys} using $\Delta t$ \cite{gikhman2007stochastic}
	\STATE Calculate $h(x_{ij})$
	\ENDFOR
	\STATE Estimate $\tilde{\A} h(x_i)$ using \eqref{E:est_Ah}
	\STATE Calculate training data using 
	\begin{equation}\label{E:data}
	    \tilde{\Delta}(x_i)=\tilde{\A} h(x_i)-(L_fh(x_i)+L_gh(x_i)u)
	\end{equation}
	\STATE Add training data into data-set, $D\leftarrow \{x_i, \tilde{\Delta}(x_i)\}$
	\ENDFOR
	\end{algorithmic}
\end{algorithm}

\section{Simulation Results}\label{Sec: sim}

\subsection{Example 1}\label{sec:example1}
\begin{figure}[htbp]
    \centering
    \includegraphics[scale=.3]{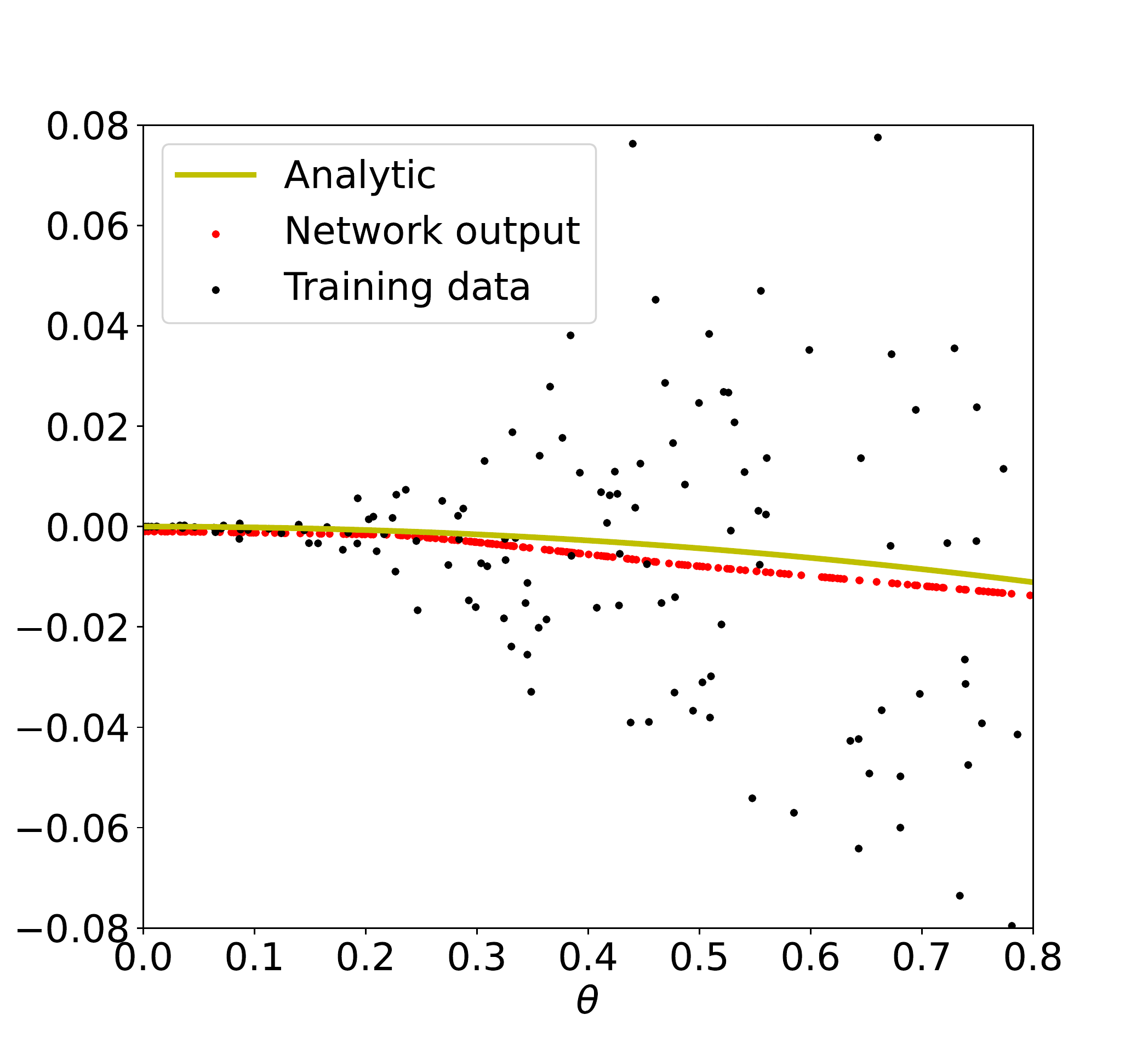}
    \caption{Training result of $\frac{1}{2}\operatorname{tr}\left[(bb^T)(x)\cdot h_{xx}(x)\right]$. The black dots are the training data. The yellow curve is the analytic result, which is the true value  $-\sqrt{3}\cdot(0.1\theta)^2$ and the red dots are the output of the neural network after training.}
    \label{fig:pendulum}
\end{figure}

In the first example, we test our result using an inverted pendulum. The system is an SDE of the form 
\begin{equation*}
d\begin{bmatrix}
\theta\\
\dot{\theta}
\end{bmatrix}=\begin{bmatrix}
\dot{\theta}\\
\frac{g}{l}\sin{\theta}
\end{bmatrix}dt+\begin{bmatrix}
0\\
\frac{1}{ml^2}
\end{bmatrix}udt+\begin{bmatrix}
0.1\theta\\
0
\end{bmatrix}dW,
\end{equation*}
with the state $x=[\theta, \dot{\theta}]^T$, gravitational acceleration $g=10$ and length $\ell=0.7$. We assume that the diffusion part $b(x)=[0.1\theta, 0]^T$ is unknown to us. Consider the control barrier function 
\begin{equation*}
    h(x)=c-x^TPx,
\end{equation*}
where
\begin{equation*}
    P=\begin{bmatrix}
        \sqrt{3}&1\\
        1&\sqrt{3}
        \end{bmatrix}.
\end{equation*}
So
\begin{equation*}
    h=0.2-\sqrt{3}\theta^2-2\theta\dot{\theta}-\sqrt{3}\dot{\theta}^2.
\end{equation*}
Accordingly, we have 
\begin{equation*}
\begin{split}
L_fh(x)+L_gh(x)u&=-(2\sqrt{3}\theta+2\dot{\theta})\dot{\theta}\\
&-(2\theta+2\sqrt{3}\dot{\theta})\cdot\frac{g}{l}\sin{\theta}-\frac{2\theta+2\sqrt{3}\dot{\theta}}{ml^2}u.
\end{split}
\end{equation*}
We follow Algorithm \ref{alg:alg} to obtain training data. We randomly sample 200 points within state space and at each point $x_i$, we simulate 50000 one-step transitions to get to the next point $x_{ij}$. The time step of the transition is $t=0.01s$. Then we estimate $\hat{\A}h(x_i)$ using \eqref{E:est_Ah}.
As a result, the training data is obtained according to \eqref{E:data}.
We use a neural network with two hidden layers, with 100 and 30 nodes for each layer, respectively, to fit the training data. We train the network with 500 epochs and compare the training result with the analytic result calculated as
\begin{equation}\label{E:pen_ana}
\begin{split}
    \frac{1}{2}\operatorname{tr}\left[(bb^T)(x)\cdot h_{xx}(x)\right]&=\frac{1}{2}tr\bigg(\begin{bmatrix}
0.1\theta & 0\\
\end{bmatrix}\\
&\cdot\begin{bmatrix}
-2\sqrt{3}& -2\\
-2& -2\sqrt{3}\\
\end{bmatrix}\cdot\begin{bmatrix}
0.1\theta \\
0\\
\end{bmatrix}\bigg)\\
&=-\sqrt{3}\cdot(0.1\theta)^2.
\end{split}
\end{equation}
The result of learning is shown in the Figure~\ref{fig:pendulum}. The black dots are the training data, the yellow curve is the analytic result calculated as in \eqref{E:pen_ana} and the red dots are the neural network output for validation after training. 

We also test the control result of applying the DDSCBF scheme. We compare the safe rate using the real SCBF, the DDSCBF scheme and CBF on the unknown system. For each case, we randomly sample 1000 trajectories and compute the safe rate. As shown in the Table~\ref{tab:pendulum}, the system is sensitive to the noise that all the trajectories are unsafe when using CBF. But after applying the DDSCBF scheme, the success rate of the system is over $90\%$.
\begin{table}[htbp]
\centering
\begin{tabular}{lcccl}
\toprule
& Success rate \\ 
\midrule
SCBF & 92\% &\\ 
DDSCBF & 91\% &\\
CBF & 0\% &\\
\bottomrule
\end{tabular}
\caption{The success rate of using SCBF, DDSCBF scheme and CBF for pendulum system over 1000 runs. }
\label{tab:pendulum}
\end{table}

\subsection{Example 2}
\begin{figure}[htbp]
    \centering
    \includegraphics[scale=.28]{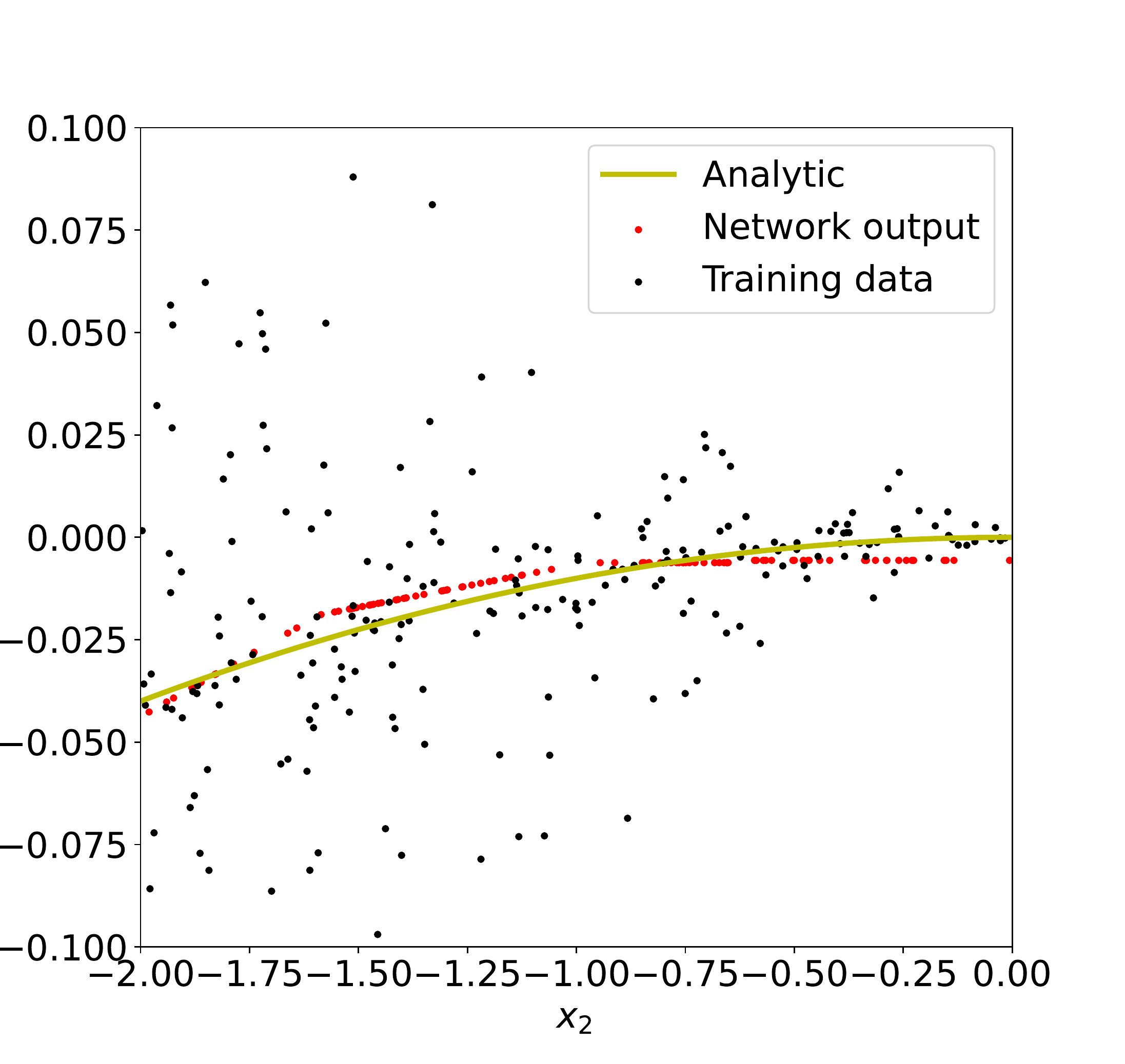}
    \caption{Training result of $\frac{1}{2}\operatorname{tr}\left[(bb^T)(x)\cdot h_{xx}(x)\right]$ for $b(x_2)=0.1x_2$. The black dots are the training data. The yellow curve is the analytic value and the red dots are the output of the neural network after training.}
    \label{fig:nonlinear_NN}
\end{figure}

In the second example, we test our result using a non-linear system given by the following stochastic differential equation:
\begin{equation*}
d\begin{bmatrix}
\dot{x}_1\\
\dot{x}_2
\end{bmatrix}=\begin{bmatrix}
-0.6x_1-x_2\\
x_1^3
\end{bmatrix}dt+\begin{bmatrix}
0\\
x_2
\end{bmatrix}udt+\begin{bmatrix}
0\\
b(x_2)
\end{bmatrix}dW.
\end{equation*}
The control objective is to reach the origin $(0,0)$ and the safe region is defined as 
\begin{equation*}
    h=-x_2^2-x_1+1>0.
\end{equation*}
The generator of $h$ is calculated as
\begin{equation*}
    \A(h)=0.6x_1+x_2-2x_1^3x_2-2x_2^2u-b(x_2)^2.
\end{equation*}
We use the same number of sample points and number of transitions at each point as in the first example. The structure of the neural network is also the same as in the first example. 
The training result is shown in the Figure~\ref{fig:nonlinear_NN}.

\begin{table}[htbp]
\centering
\begin{tabular}{lcccl}
\toprule
& $b(x_2)=0.1x_2$ & $b(x_2)=0.15x_2$ \\ 
\midrule
SCBF & 86.8\% & 84.3\% &\\ 
DDSCBF & 85.2\% & 83\% &\\
CBF & 77.3\% & 57.5\% & \\
\bottomrule
\end{tabular}
\caption{The success rate of using SCBF, learned SCBF and CBF for non-linear system over 1000 runs under different noise with $b(x_2)=0.1x_2$ and $b(x_2)=0.15x_2$. }
\label{tab:nonlinear}
\end{table}

\begin{figure}[htbp]
	\centering
	\begin{subfigure}[h]{0.48\linewidth}
		\includegraphics[width=1\linewidth]{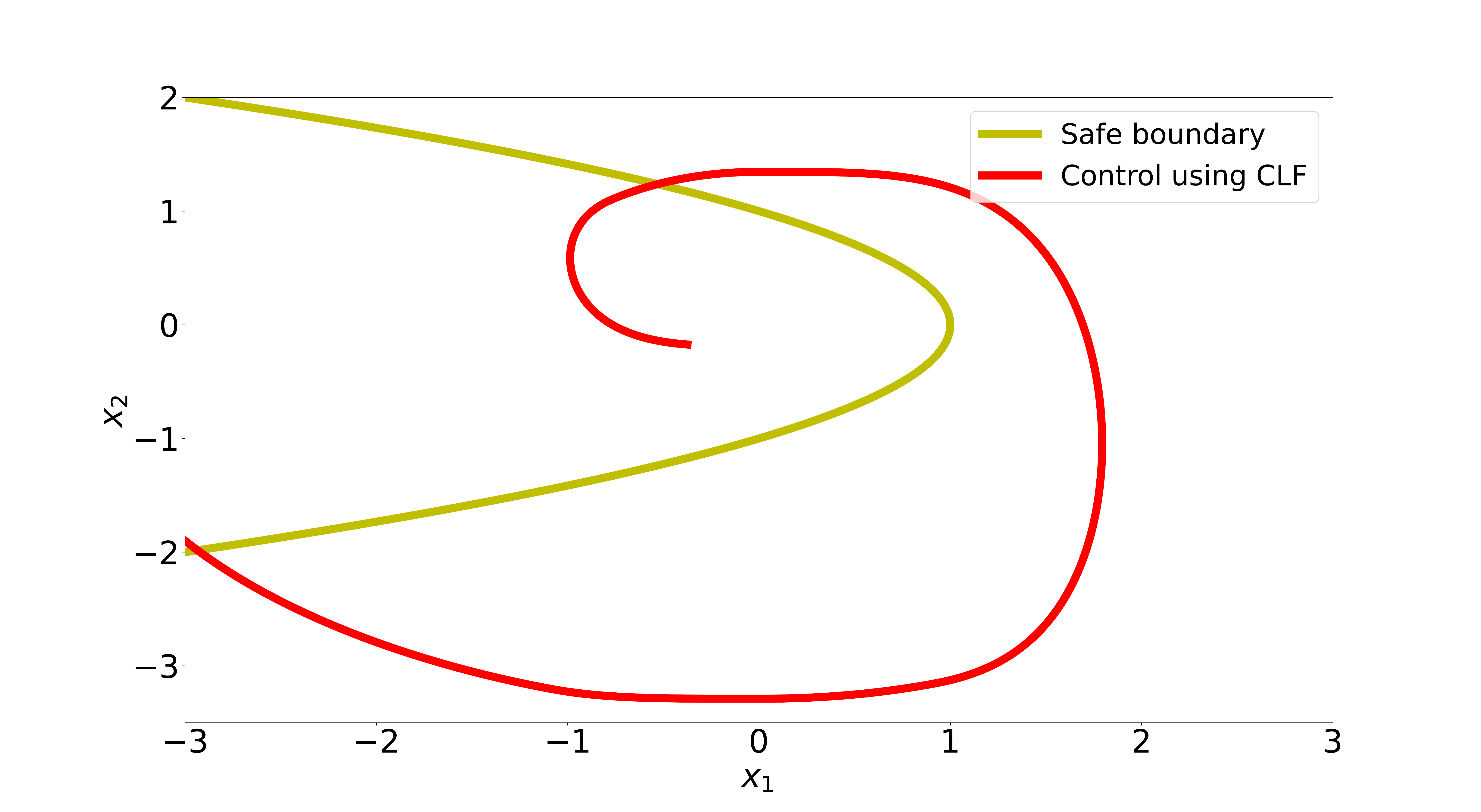}
	\caption{}
	\label{fig:nonlinear_clf}
	\end{subfigure}%
	\begin{subfigure}[h]{0.48\linewidth}
		\includegraphics[width=1\linewidth]{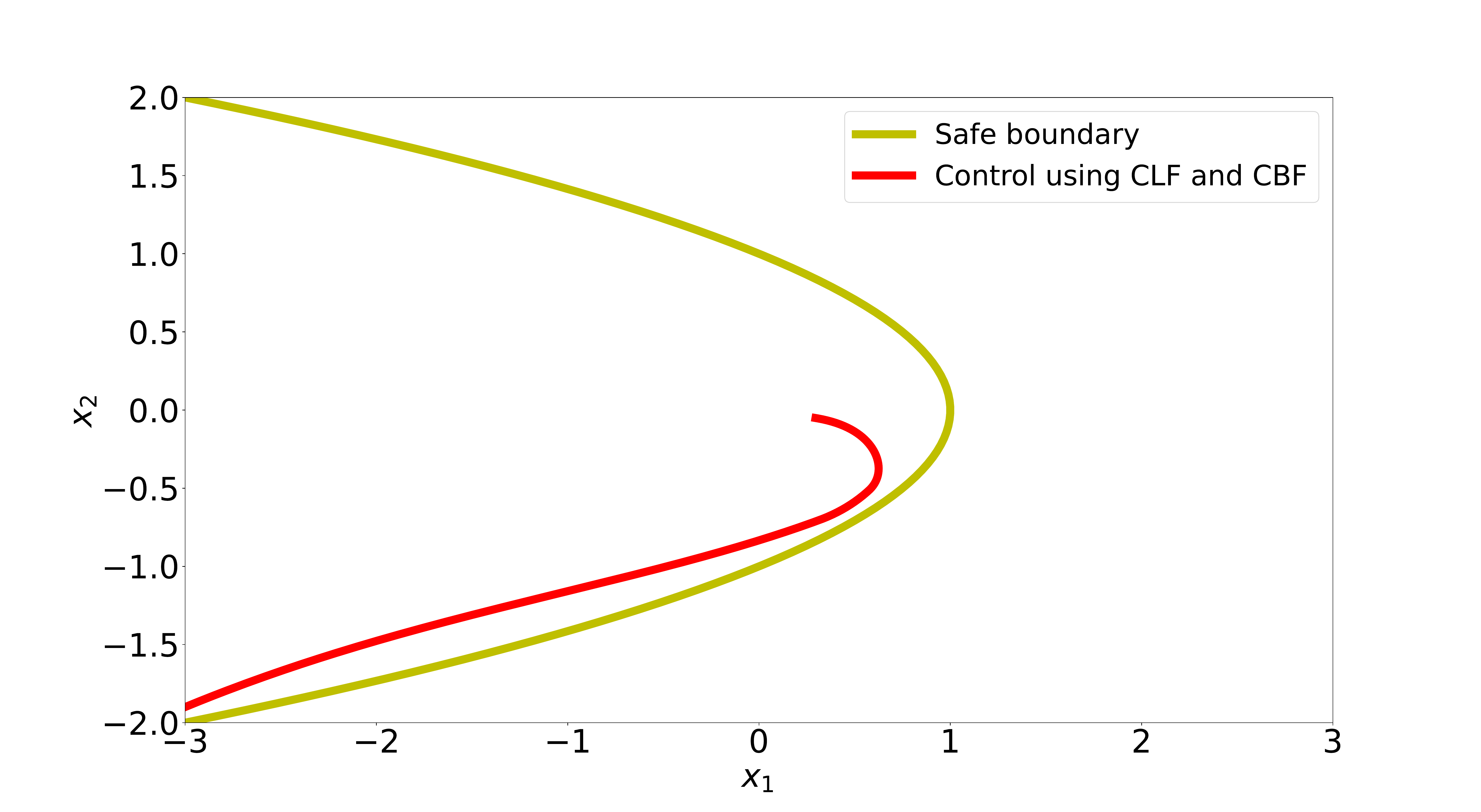}
	\caption{}
	\label{fig:nonlinear_cbf}
	\end{subfigure}
	\begin{subfigure}[h]{0.48\linewidth}
		\includegraphics[width=1\linewidth]{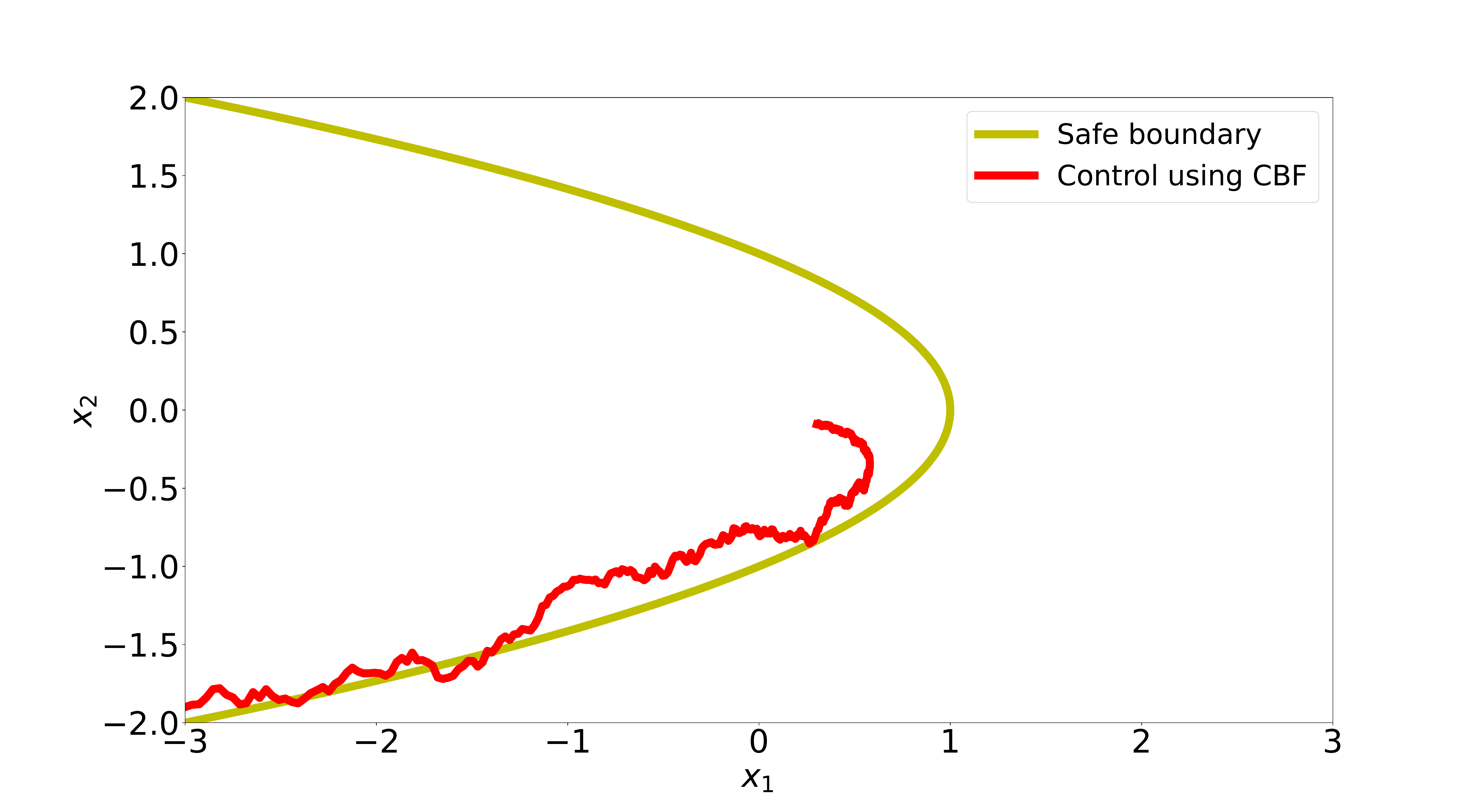}
	\caption{}
	\label{fig:nonlinear_noise_cbf}
	\end{subfigure}%
	\begin{subfigure}[h]{0.48\linewidth}
		\includegraphics[width=1\linewidth]{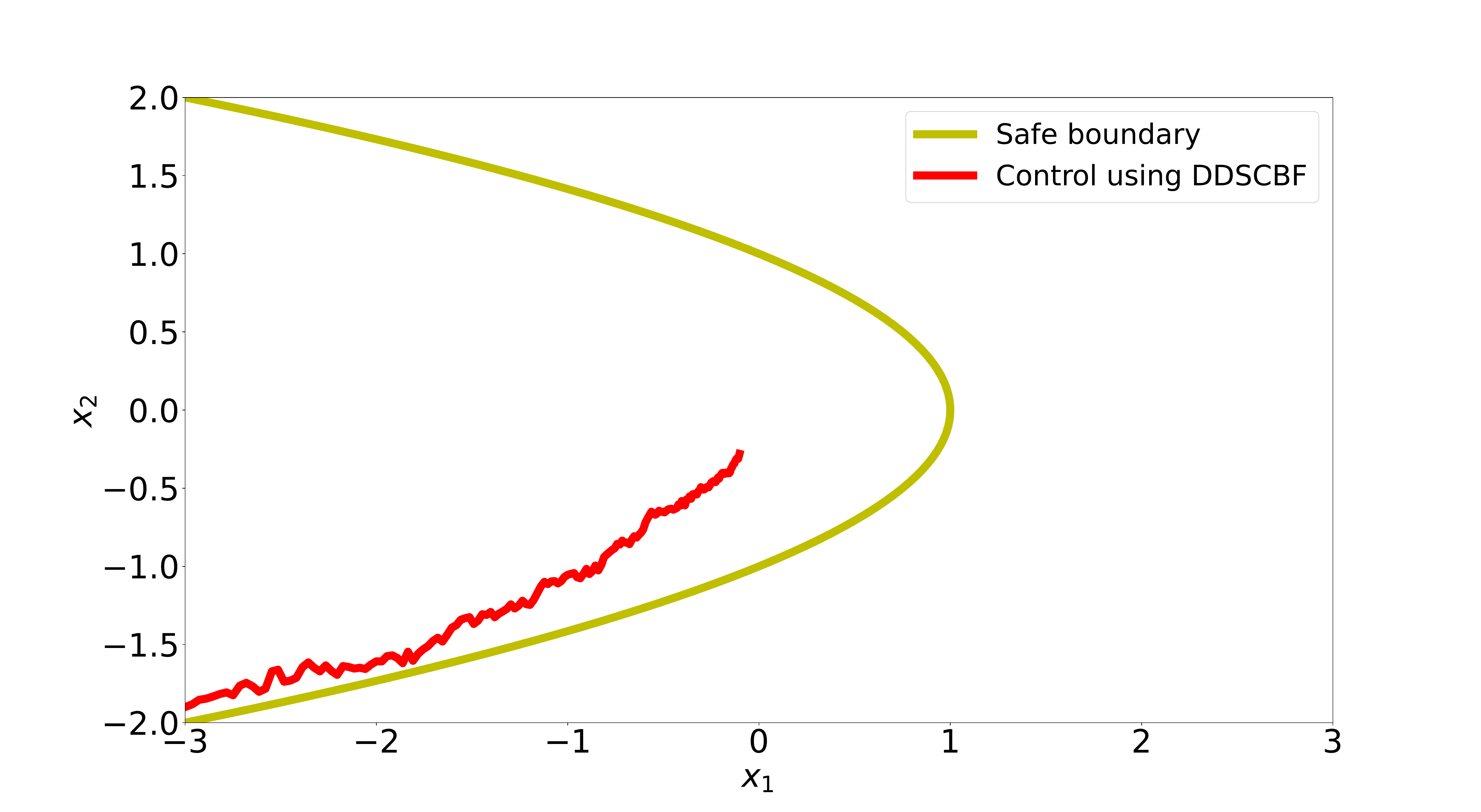}
	\caption{}
	\label{fig:nonlinear_noise_scbf}
	\end{subfigure}
	\caption{Simulation result of Example 2. (a): Control of system using CLF with $b(x_2)=0$. (b): Control of deterministic system using CLF and CBF with $b(x_2)=0$ . (c): Sample trajectory of uncertain system using CBF with $b(x_2)=0.1x_2$. (d): Sample trajectory of uncertain system using DDSCBF scheme with $b(x_2)=0.1x_2$.}
\end{figure}

We use a CLF to control the deterministic system, i.e., $\b(x_2)=0$ and the result is shown in Figure~\ref{fig:nonlinear_clf}. Also the control using CLF and CBF for the deterministic system is shown in Figure~\ref{fig:nonlinear_cbf}. We can see that CBF will guarantee a safe trajectory for the deterministic system. However, when the system has a diffusion part of $b(x_2)=0.1x_2$, the noise will make the trajectory unsafe using CBF as shown in Figure~\ref{fig:nonlinear_noise_cbf}. By using our DDSCBF scheme, the trajectory is within the safe region as shown in Figure~\ref{fig:nonlinear_noise_scbf}.

As in the first example, in order to test the performance of our DDSCBF scheme, we randomly sample 1000 trajectories and compute the safe rate under different noise for $b(x_2)=0.1x_2$ and $b(x_2)=0.15x_2$. The result is presented in Table~\ref{tab:nonlinear}.

\section{Conclusion}
In this work, we study safety-critical control of stochastic systems with unknown diffusion parts. We use supervised learning to approximate the derivative of SCBF for safety control of SDEs and show that our DDSCBF scheme can approximate the derivative of SCBF with guarantee. We also validate our result using two non-linear SDEs. Extension of this work in the future will be mainly in two aspects. The first one is to study uncertainty in not only the diffusion part, but also in the drift part. We will also compare results using different learning method such as Gaussian regression. Also, the computation efficiency of DDSCBF scheme is much lower when applied to systems with higher relative degree. So the second direction in the future will be focused on SDEs with unknown part with high relative degree. 
\bibliographystyle{plain}
\bibliography{root}{}

\begin{thebibliography}{10}

\bibitem{ames2019control}
Aaron~D Ames, Samuel Coogan, Magnus Egerstedt, Gennaro Notomista, Koushil
  Sreenath, and Paulo Tabuada.
\newblock Control barrier functions: Theory and applications.
\newblock In {\em Proc. of ECC}, pages 3420--3431. IEEE, 2019.

\bibitem{ames2014control}
Aaron~D Ames, Jessy~W Grizzle, and Paulo Tabuada.
\newblock Control barrier function based quadratic programs with application to
  adaptive cruise control.
\newblock In {\em Proc. of CDC}, pages 6271--6278. IEEE, 2014.

\bibitem{ames2016control}
Aaron~D Ames, Xiangru Xu, Jessy~W Grizzle, and Paulo Tabuada.
\newblock Control barrier function based quadratic programs for safety critical
  systems.
\newblock {\em IEEE Transactions on Automatic Control}, 62(8):3861--3876, 2016.

\bibitem{barlow1988brownian}
Martin~T Barlow and Edwin~A Perkins.
\newblock Brownian motion on the sierpinski gasket.
\newblock {\em Probability theory and related fields}, 79(4):543--623, 1988.

\bibitem{chen2017obstacle}
Yuxiao Chen, Huei Peng, and Jessy Grizzle.
\newblock Obstacle avoidance for low-speed autonomous vehicles with barrier
  function.
\newblock {\em IEEE Transactions on Control Systems Technology},
  26(1):194--206, 2017.

\bibitem{clark2019control}
Andrew Clark.
\newblock Control barrier functions for complete and incomplete information
  stochastic systems.
\newblock In {\em Proc. of ACC}, pages 2928--2935. IEEE, 2019.

\bibitem{garcia2015comprehensive}
Javier Garc{\i}a and Fernando Fern{\'a}ndez.
\newblock A comprehensive survey on safe reinforcement learning.
\newblock {\em Journal of Machine Learning Research}, 16(1):1437--1480, 2015.

\bibitem{gikhman2007stochastic}
Iosif~Ilyich Gikhman and Anatoli~Vladimirovich Skorokhod.
\newblock Stochastic differential equations.
\newblock In {\em The theory of stochastic processes III}, pages 113--219.
  Springer, 2007.

\bibitem{girard2006efficient}
Antoine Girard, Colas Le~Guernic, and Oded Maler.
\newblock Efficient computation of reachable sets of linear time-invariant
  systems with inputs.
\newblock In {\em Proc. of HSCC}, pages 257--271. Springer, 2006.

\bibitem{hsu2015control}
Shao-Chen Hsu, Xiangru Xu, and Aaron~D Ames.
\newblock Control barrier function based quadratic programs with application to
  bipedal robotic walking.
\newblock In {\em Proc. of ACC}, pages 4542--4548. IEEE, 2015.

\bibitem{kisielewicz2007stochastic}
Micha{\l} Kisielewicz.
\newblock Stochastic differential inclusions and diffusion processes.
\newblock {\em Journal of mathematical analysis and applications},
  334(2):1039--1054, 2007.

\bibitem{kisielewicz2013stochastic}
Micha{\l} Kisielewicz et~al.
\newblock {\em Stochastic differential inclusions and applications}.
\newblock Springer, 2013.

\bibitem{kushner1967stochastic}
Harold~J Kushner.
\newblock {\em Stochastic Stability and Control}.
\newblock Academic Press, 1967.

\bibitem{lamport1977proving}
Leslie Lamport.
\newblock Proving the correctness of multiprocess programs.
\newblock {\em IEEE Transactions on Software Engineering}, (2):125--143, 1977.

\bibitem{meng2022acc}
Yiming Meng and Jun Liu.
\newblock Sufficient conditions for robust probabilistic reach-avoid-stay
  specifications using stochastic lyapunov-barrier functions.
\newblock In {\em In Proceedings of 2022 American Control Conference (ACC)}.

\bibitem{nejati2021data}
Ameneh Nejati, Abolfazl Lavaei, Sadegh Soudjani, and Majid Zamani.
\newblock Data-driven estimation of infinitesimal generators of stochastic
  systems.
\newblock {\em IFAC-PapersOnLine}, 54(5):277--282, 2021.

\bibitem{nguyen2016exponential}
Quan Nguyen and Koushil Sreenath.
\newblock Exponential control barrier functions for enforcing high
  relative-degree safety-critical constraints.
\newblock In {\em Proc. of ACC}, pages 322--328. IEEE, 2016.

\bibitem{oksendal2003stochastic}
Bernt {\O}ksendal.
\newblock Stochastic differential equations.
\newblock In {\em Stochastic differential equations}, pages 65--84. Springer,
  2003.

\bibitem{ratschan2007safety}
Stefan Ratschan and Zhikun She.
\newblock Safety verification of hybrid systems by constraint propagation-based
  abstraction refinement.
\newblock {\em ACM Transactions on Embedded Computing Systems (TECS)},
  6(1):8--es, 2007.

\bibitem{rauscher2016constrained}
Manuel Rauscher, Melanie Kimmel, and Sandra Hirche.
\newblock Constrained robot control using control barrier functions.
\newblock In {\em Proc. of IROS}, pages 279--285. IEEE, 2016.

\bibitem{santoyo2021barrier}
Cesar Santoyo, Maxence Dutreix, and Samuel Coogan.
\newblock A barrier function approach to finite-time stochastic system
  verification and control.
\newblock {\em Automatica}, 125:109439, 2021.

\bibitem{sarkar2020high}
Meenakshi Sarkar, Debasish Ghose, and Evangelos~A Theodorou.
\newblock High-relative degree stochastic control {L}yapunov and barrier
  functions.
\newblock {\em arXiv preprint arXiv:2004.03856}, 2020.

\bibitem{taylor2020learning}
Andrew Taylor, Andrew Singletary, Yisong Yue, and Aaron Ames.
\newblock Learning for safety-critical control with control barrier functions.
\newblock In {\em Proc. of L4DC}, pages 708--717. PMLR, 2020.

\bibitem{wang2021learning}
Chuanzheng Wang, Yiming Meng, Yinan Li, Stephen~L Smith, and Jun Liu.
\newblock Learning control barrier functions with high relative degree for
  safety-critical control.
\newblock In {\em 2021 European Control Conference (ECC)}, pages 1459--1464.
  IEEE, 2021.

\bibitem{wang2021safety}
Chuanzheng Wang, Yiming Meng, Stephen~L Smith, and Jun Liu.
\newblock Safety-critical control of stochastic systems using stochastic
  control barrier functions.
\newblock In {\em 2021 IEEE Conference on Decision and Control (CDC)}, pages
  5924--5931, 2021.

\bibitem{xiao2019control}
Wei Xiao and Calin Belta.
\newblock Control barrier functions for systems with high relative degree.
\newblock In {\em Proc. of CDC}, pages 474--479. IEEE, 2019.

\end{thebibliography}

\appendix
We prove the $L_1$ convergence of LLN as in \eqref{E: l1}. We first introduce the following theorem as we will use in the final proof. 
\begin{definitionA}[Backward martingale]
A backward martingale is a stochastic process $\{X_{-n}\}_{n=1,2,\cdots}$ such that, for each $n$, $X_{-n}$ is $L_1$ integrable and $\mathscr{F}_{-n}$-measurable, and satisfies
\begin{equation}
    \ee[X_{-n-1}\;|\;\mathscr{F}_{-n}]=X_{-n}.
\end{equation}
\end{definitionA}

\begin{thmA}[Backward martingale convergence theorem]
For every backward maringale, as $n\rightarrow \infty$,
\begin{equation}
    X_{-n}\rightarrow \ee[X_{-1}\;|\;\mathscr{F}_{-\infty}]\;\;\pp\text{-a.s. and in}\; L_1. 
\end{equation}
\end{thmA}
\begin{thmA}[Kolmogorov's $0$-$1$-law]
Let $\mathscr{F}_1,\mathscr{F}_2,\cdots$ be independent $\sigma$-fields and denote by $\mathscr{F}_{\infty}=\cap_{n=1}^\infty\sigma\left(\cup_{k=n}^\infty \mathscr{F}_k\right)$ the corresponding tail field. Then 
$$\pp[A]\in\{0,1\},\;\;\forall A\in\mathscr{F}_\infty.$$
\end{thmA}
\noindent\textbf{Proof of \eqref{E: l1}}: Let $Y_i=X^{u,(i)}_{\tau_s}$ for $i\in\{1,2,\cdots\}$.
Then $\{Y_i\}$ is $L_1$ integrable and i.i.d. w.r.t. $\pp^x$. Let $S_n=\sum_{i=1}^nY_i$ be the finite sum and let $X_{-n}=\frac{S_n}{n}$ be the average.  Then 
the $\sigma$-field $\mathscr{F}_{-n}=\sigma\{S_n,S_{n+1},\cdots\}$ is a decreasing filtration. Due to the independence of $\{Y_i\}$, we have
\begin{equation}\label{E: backward}
\begin{split}
       \ee^x[X_{-1}\;|\;\mathscr{F}_{-n}]& =\ee^x[Y_1\;|\;S_n,S_{n+1},\cdots]\\
       &=\ee^x[Y_1\;|\;S_n,Y_{n+1},Y_{n+2},\cdots]\\
       & =\ee^x[Y_1\;|\;S_n]. \end{split}
\end{equation}
 Notice that  $\ee^x[Y_i\;|\;S_n]=\ee^x[Y_j\;|\;S_n]$  by symmetry for $i,j\in\{1,\cdots,n\}$,
  then 
 \begin{equation}
     n\ee^x[Y_i\;|\;S_n]=\sum_{i=1}^n\ee^x[Y_i\;|\;S_n]=\ee^x[S_n\;|\;S_n]=S_n.
 \end{equation}
Combining the above, we have
$$\ee^x[X_{-1}\;|\;\mathscr{F}_{-n}]=\frac{S_n}{n}=X_{-n}, $$
which verifies that $\{X_{-n}\}$ is a backward martingale. By the backward martingale convergence theorem, we immediately have
$$\frac{S_n}{n}\rightarrow \ee^x[Y_1\;|\;\mathscr{F}_{-\infty}],\;\;\pp^x\text{-a.s. and in}\; L_1. $$
By Kolmogorov's $0$-$1$ law, we have that all $A$ in the tail field $ \mathscr{F}_{-\infty}$ have probability either $0$ or $1$, which in turn implies that the conditional expectation $\ee^x[Y_1\;|\;\mathscr{F}_{-\infty}]$ must be a constant (by the definition of conditional expectation) and should be equal to the average $\ee^x[Y_1]=\eee^x[Y_1]$.
\end{document}